\documentclass[11pt,a4paper]{article}
\usepackage{epsfig,graphicx}
\usepackage[margin=3.5cm]{geometry}

\usepackage[sort&compress,numbers]{natbib}
\usepackage{doi}
\usepackage{hyperref}

\usepackage[usenames,dvipsnames,svgnames,table]{xcolor}  
\hypersetup{colorlinks=true, urlcolor=Cerulean, citecolor=Cerulean}   

\usepackage{url}
\usepackage[utf8]{inputenc} 
\usepackage{amssymb} 
\usepackage{amsmath} 
\usepackage{braket} 
\usepackage{slashed} 
\usepackage{multirow} 
\usepackage{soul} 
\usepackage{color} 
\usepackage{booktabs} 
\usepackage[font=footnotesize,labelfont=bf]{caption}

\title{Correlated low-energy constants in large-$N_c$ chiral perturbation theory}
\author{Pere Masjuan$^{1,2}$\thanks{masjuan@ifae.es} }

\date{
	$^1$Grup de Física Teòrica, Departament de Física, 
Universitat Autònoma de Barcelona.
\\
	$^2$Institut de Física d’Altes Energies (IFAE) and 
The Barcelona Institute of Science and Technology (BIST), 
Campus UAB, 08193 Bellaterra (Barcelona), Spain \\ 
}

\begin{document}
	\maketitle
	
\begin{abstract}

The combined chiral and large-$N_c$ expansion is increasingly employed in precision studies involving the $\eta$ and $\eta'$ mesons, including recent applications to low-energy axion phenomenology within U(3) chiral perturbation theory. We point out that the operator structure of the large-$N_c$ chiral Lagrangian naturally induces correlated directions among the low-energy constants $(F_0,L_4,C_{16})$ and $(F_0,L_6,C_{20})$, implying that phenomenological analyses determine correlated combinations of couplings rather than independent low-energy constants. Using the determination of the pion decay constant as an illustrative example, we reinterpret existing phenomenological and lattice determinations in terms of these correlated directions, showing that the well-known anticorrelation between $F_0$ and $L_4$ extends naturally to NNLO through $C_{16}$. These correlated directions lead to a practical prescription for interpreting and propagating phenomenological determinations of the low-energy constants consistently within the combined chiral and large-$N_c$ framework.

\end{abstract}

\section{Introduction}

Chiral Perturbation Theory (ChPT) is the effective field theory of QCD at low energies and provides the standard framework for describing the interactions of the light pseudoscalar mesons \cite{Gasser:1983yg,Gasser:1984gg}. Continuous experimental progress, together with increasingly precise lattice-QCD calculations, has led to progressively more accurate determinations of the low-energy constants (LECs) entering the chiral Lagrangian \cite{Aoki:2013ldr,FlavourLatticeAveragingGroupFLAG:2024oxs}. A comprehensive assessment of these determinations was presented in Ref.~\cite{Bijnens:2014lea}, whose recommended values continue to be widely adopted in phenomenological applications.

Besides its traditional role in meson phenomenology, ChPT has recently acquired renewed importance through searches for axions and axion-like particles. In particular, the description of axion couplings to the light pseudoscalar mesons \cite{GrillidiCortona:2015jxo,Landini:2019eck,DiLuzio:2022gsc,DiLuzio:2022tbb,Wang:2024tre,Meggiolaro:2025yiu,Alda:2025uwo}, axion-baryon coupling \cite{Bigazzi:2019hav,Vonk:2021sit,Cao:2024cym}, axion–meson mixing \cite{Gao:2022xqz}, rare $\eta$ and $\eta'$ decays \cite{Gan:2020aco}, and axion-photon couplings \cite{Gao:2024vkw,Feruglio:2025xvc} relies on the U(3) framework making the determination of the corresponding LECs increasingly relevant for present and future experimental searches.

These developments have renewed interest in the determination and
interpretation of the corresponding low-energy constants within the
combined chiral and large-$N_c$ expansion, including studies of the
range of validity of the low-energy effective theory itself
\cite{DiLuzio:2023cuk}. They are becoming increasingly relevant for ongoing and future experimental programmes searching for light weakly interacting particles and rare meson decays \cite{Goudzovski:2022vbt}, including reference searches at REDTOP \cite{REDTOP:2022slw,REDTOP:2026joh}, NA62 \cite{NA62:2021zjw}, NA64 \cite{NA64:2020qwq}, KOTO \cite{KOTO:2020prk}, BESIII \cite{BESIII:2024hdv,BESIII:2024awu,Zhang:2026pyv}, HADES \cite{Zielinski:2026lqa}, BELLE-II \cite{Belle-II:2020jti}, MicroBooNE \cite{MicroBooNE:2021sov}, BABAR \cite{BaBar:2021ich}, among others.

Global fits determine the combinations of LECs constrained by physical observables \cite{Bijnens:2014lea}. Nevertheless, the fitted values are often employed as if the individual LECs were independently determined. However, the operator structure of the large-$N_c$ chiral Lagrangian \cite{Leutwyler:1996sa,Herrera-Siklody:1996tqr,Kaiser:2000gs} naturally induces strong correlations among several of these couplings \cite{Ecker:2010nc,Ecker:2013pba}. Such correlations are already reflected in phenomenological analyses, although their interpretation has received comparatively little attention. In particular, the leading-order decay constant $F_0$ is strongly correlated with the next-to-leading-order couplings $L_4$ and $L_6$, and these correlations extend naturally to the next order through the couplings $C_{16}$ and $C_{20}$ \cite{Ecker:2010nc,Ecker:2013pba}. Consequently, phenomenological fits determine correlated directions in the multidimensional parameter space of the LECs rather than isolated values of the individual couplings. 
This distinction becomes particularly relevant as phenomenological applications reach a level of precision where the uncertainty associated with these correlated directions may compete with other subleading effects, such as isospin-breaking corrections.

In this work we identify the correlated directions implied by the operator structure of the large-$N_c$ chiral Lagrangian and discuss their implications for the interpretation and use of phenomenological determinations of the LECs. We show that the large-$N_c$ organization of the chiral Lagrangian naturally identifies two correlated families of couplings up to NNLO,
\[
(F_0,L_4,C_{16}),\qquad
(F_0,L_6,C_{20}),
\]
which are constrained through specific combinations entering physical observables. This observation provides a simple interpretation of the correlations found in phenomenological fits and suggests that whenever the combined chiral and large-$N_c$ expansion is adopted—as in U(3) ChPT—the choice among phenomenologically equivalent solutions should be consistent with the underlying large-$N_c$ operator hierarchy. The discussion is illustrated using the determinations of Ref.~\cite{Bijnens:2014lea}. An analogous correlation is expected for the operator chain $(F_0,L_6,C_{20})$, although we do not explore it explicitly here.

\section{Correlated directions in the LEC parameter space}

The correlations discussed in this work originate directly from the operator structure of the large-$N_c$ chiral Lagrangian. In the combined chiral and large-$N_c$ expansion, the relevant kinetic operator receives successive contributions at LO, NLO and NNLO proportional to the couplings $(F_0,L_4,C_{16})$,

\begin{equation}
\frac{1}{4}\langle u_\mu u^\mu \rangle
\left[
F_0^2
+4L_4\langle\chi_+\rangle
+4C_{16}\langle\chi_+^2\rangle
+\cdots
\right].
\end{equation}

In the decay-constant sector considered here, physical observables are
therefore sensitive to this combination rather than to the individual
LECs separately. The natural object determined by phenomenology is thus
a correlated direction in the $(F_0,L_4,C_{16})$ parameter space. Numerically, since $\langle \chi_+ \rangle \sim (4M_K^2)$, the relevant combination is approximately

\begin{equation}\label{eq2}
F^2_{\pi} \sim F^2_0 + 16 L_4 M_K^2 + 64 C_{16} M_K^4 + \cdots ,
\end{equation}

The correlated operator chain does not terminate at NNLO. Higher-order
operators containing additional powers of $\chi_+$ generate analogous
contributions involving further low-energy constants. Schematically, the
decay constant therefore receives a tower of terms of the form
\[
F_\pi^2 = F_0^2+16L_4 M_K^2+64C_{16}M_K^4+D_i\,M_K^6+E_i\,M_K^8+\cdots ,
\]
where the precise operator combinations and their coefficients beyond
NNLO are not presently constrained by the analysis considered here.
The truncation at NNLO in Eq.~(\ref{eq2}) should therefore be understood as a limitation of the available phenomenological information, rather than as a
termination of the underlying operator correlation.

The predictive use of the expansion requires the resulting tower to exhibit a numerical hierarchy, which may arise from decreasing higher-order coefficients or, more generally, from cancellations among successive contributions. The large numerical scale associated with $M_K^2$ (since $4M_K^2 \sim 1\,\mathrm{GeV}^2$), however, makes such convergence a nontrivial assumption in the strange-meson sector.

\begin{figure}
    \centering
    \includegraphics[width=0.45\linewidth]{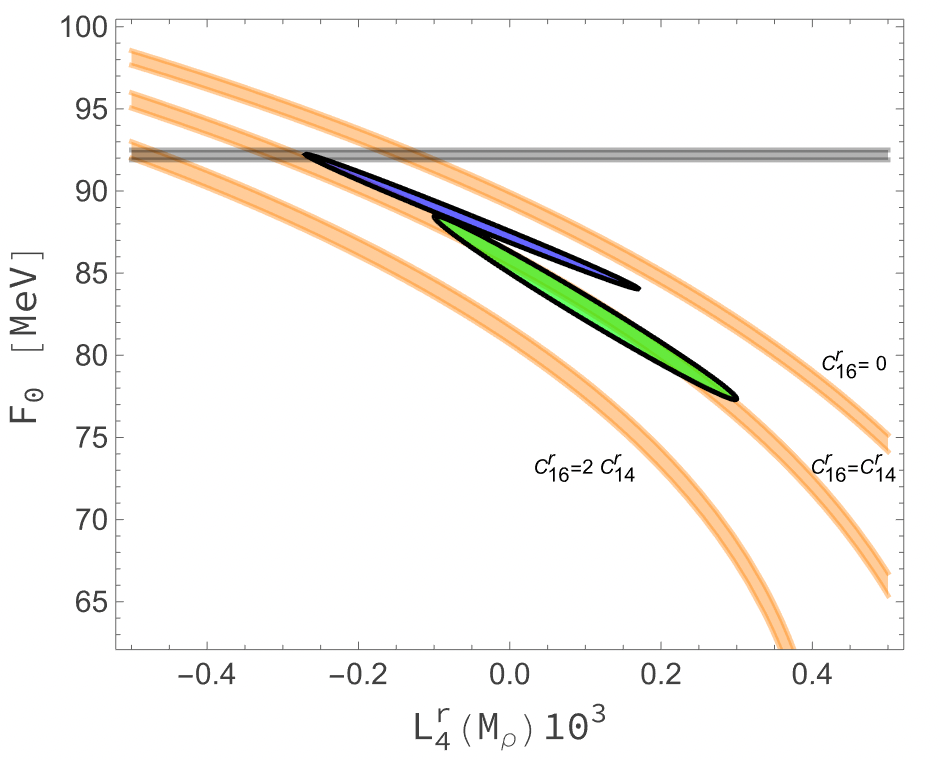}
    \includegraphics[width=0.45\linewidth]{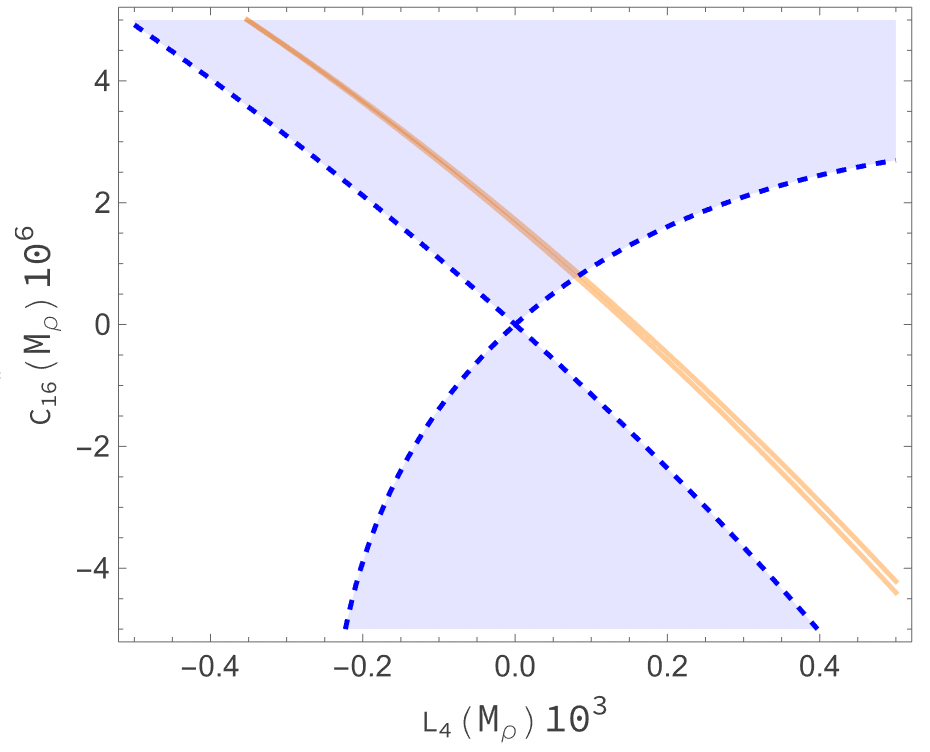}
    \caption{
Left: Correlation between $F_0$ and $L_4$ obtained from the NNLO analysis of Ref.~\cite{Ecker:2013pba}. The blue and green ellipses denote the RBC/UKQCD lattice constraints \cite{RBC:2010qam,RBC:2012cbl}. The orange bands correspond to three representative values of $C_{16}$, illustrating the correlated operator chain $(F_0,L_4,C_{16})$. Right: Correlated values of $L_4$ and $C_{16}$ for fixed $F_0=86.0(5)\,$MeV. The shaded regions indicate where the NNLO contribution exceeds the NLO one.
}\label{fig1}
\end{figure}

Figure~\ref{fig1} illustrates the resulting correlation geometry in the $(F_0,L_4,C_{16})$ parameter space. The left panel explains why precise determinations of $F_0$ remain difficult. Lattice and phenomenological information constrain primarily a correlated combination of $F_0$ and $L_4$, while at NNLO this degeneracy naturally extends to $C_{16}$ through Eq.~(\ref{eq2}). Consequently, these couplings should not be interpreted as independently determined quantities.

The right panel shows the complementary projection of the same correlated family in the $(L_4,C_{16})$ plane for fixed $F_0=86.0(5)$ MeV. The shaded regions indicate where the NNLO contribution becomes larger than the NLO one, signalling a loss of the expected hierarchy of the truncated chiral expansion.

For the numerical illustration we adopt the leading-$N_c$ couplings $L_5$, $C_{14}$ and $C_{17}$ from the large-$N_c$-motivated analysis of Ref.~\cite{Ecker:2013pba}, since they are extracted from the same analysis of $F_K/F_\pi$ considered here. In the absence of an updated dedicated large-$N_c$ determination of the NNLO couplings, these values provide a consistent benchmark for illustrating the correlated operator structure discussed in this work.

More generally, we advocate selecting LEC determinations according to their consistency with the expected large-$N_c$ hierarchy rather than according to a particular global fit. In practice, this amounts to adopting the leading large-$N_c$ couplings from analyses specifically designed within that framework, while interpreting the OZI-suppressed couplings $L_4$, $L_6$, $C_{16}$ and $C_{20}$ as correlated quantities rather than independently determined parameters. Existing phenomenological analyses, including Ref.~\cite{Bijnens:2014lea}, should therefore be viewed as constraining correlated directions in parameter space instead of isolated values of these couplings.

The purpose of the present analysis is therefore not to provide a new determination of the LECs, but to reinterpret existing phenomenological and lattice determinations within the combined chiral and large-$N_c$ expansion. Rather than fixing OZI-suppressed couplings to isolated best-fit values, the associated uncertainties should be propagated along the correlated directions implied by the effective Lagrangian. Although we have focused on the decay-constant sector, the same reasoning applies to the second operator chain $(F_0,L_6,C_{20})$.

\section{Conclusions}

We have shown that the operator structure of the combined chiral and large-$N_c$ expansion naturally organizes the LECs into correlated families, exemplified by the chains $(F_0,L_4,C_{16})$ and $(F_0,L_6,C_{20})$. Consequently, phenomenological analyses determine correlated directions in the parameter space of the LECs rather than independent values of the individual couplings.

Using the determination of the pion decay constant as an illustrative example, we have shown that the familiar anticorrelation between $F_0$ and $L_4$ extends naturally to NNLO through $C_{16}$. This provides a simple interpretation of existing phenomenological and lattice determinations and explains the longstanding difficulty in extracting a precise value of $F_0$.

The NNLO analysis should consequently be regarded as the first
nontrivial test of this numerical hierarchy rather than as the endpoint of the correlated operator chain; higher-order terms provide an additional
source of theoretical uncertainty whenever the chiral expansion is pushed into the strange-meson sector.

The purpose of the present work is not to advocate a new global determination of the LECs, but to provide a consistent prescription for interpreting and propagating existing determinations within the combined chiral and large-$N_c$ framework. In practical applications, we advocate selecting determinations that respect the expected large-$N_c$ operator hierarchy, adopting the leading large-$N_c$ couplings from analyses performed within that framework, and propagating the uncertainty associated with OZI-suppressed couplings along the correlated directions identified here rather than treating them as independently determined quantities.

As phenomenological applications continue to improve in precision, particularly in U(3) chiral perturbation theory and low-energy axion phenomenology, these correlated directions should be regarded as an intrinsic theoretical uncertainty of the effective theory and propagated consistently in future analyses.

\section*{Acknowledgments}

This work has been supported by the Ministerio de Ciencia e Innovación under grant PID2020-112965GB-I00, by the Secretaria d’Universitats i Recerca del Departament d’Empresa i Coneixement de la Generalitat de Catalunya under grant 2021 SGR 00649, and by the Spanish Ministry of Science and Innovation (MICINN) through the State Research Agency under the Severo Ochoa Centres of Excellence Programme 2025–2029 (CEX2024-001441-S). IFAE is partially funded by the CERCA program of the Generalitat de Catalunya.



\begin{thebibliography}{99}

\bibitem{Gasser:1983yg}
J.~Gasser and H.~Leutwyler,
``Chiral Perturbation Theory to One Loop,''
Annals Phys. \textbf{158}, 142 (1984).

\bibitem{Gasser:1984gg}
J.~Gasser and H.~Leutwyler,
``Chiral Perturbation Theory: Expansions in the Mass of the Strange Quark,''
Nucl. Phys. B \textbf{250}, 465--516 (1985).

\bibitem{Aoki:2013ldr}
S.~Aoki et al.,
``Review of Lattice Results Concerning Low-Energy Particle Physics,''
Eur. Phys. J. C \textbf{74}, 2890 (2014).

\bibitem{FlavourLatticeAveragingGroupFLAG:2024oxs}
Y.~Aoki et al.,
``FLAG review 2024,''
Phys. Rev. D \textbf{113}, 014508 (2026).

\bibitem{Bijnens:2014lea}
J.~Bijnens and G.~Ecker,
``Mesonic low-energy constants,''
Ann. Rev. Nucl. Part. Sci. \textbf{64}, 149--174 (2014).

\bibitem{GrilliDiCortona:2015jxo}
G.~Grilli di Cortona, E.~Hardy, J.~Pardo Vega, and G.~Villadoro,
``The QCD axion, precisely,''
JHEP \textbf{01}, 034 (2016).

\bibitem{Landini:2019eck}
G.~Landini and E.~Meggiolaro,
``Study of the interactions of the axion with mesons and photons using a chiral effective Lagrangian model,''
Eur. Phys. J. C \textbf{80}, 302 (2020).

\bibitem{DiLuzio:2022gsc}
L.~Di Luzio, J.~Martin Camalich, G.~Martinelli, J.~A.~Oller, and G.~Piazza,
``Axion-pion thermalization rate in unitarized NLO chiral perturbation theory,'' Phys. Rev. D \textbf{108}, 035025 (2023).

\bibitem{DiLuzio:2022tbb}
L.~Di Luzio and G.~Piazza,
``$a\to\pi\pi\pi$ decay at next-to-leading order in chiral perturbation theory,'' JHEP \textbf{12}, 041 (2022).
[Erratum: JHEP \textbf{05}, 018 (2023)].

\bibitem{Wang:2024tre}
J.-B.~Wang, Z.-H.~Guo, Z.~Lu, and H.-Q.~Zhou,
``Axion production in the $\eta\to\pi\pi a$ decay within SU(3) chiral perturbation theory,''
JHEP \textbf{11}, 029 (2024).

\bibitem{Meggiolaro:2025yiu}
E.~Meggiolaro and M.~Tamburini,
``New study of the interactions of the axion with mesons and photons using a chiral effective Lagrangian model,''
Phys. Rev. D \textbf{111}, 095024 (2025).

\bibitem{Alda:2025uwo}
J.~Alda, M.~Fuentes Zamoro, L.~Merlo, X.~Ponce D\'iaz, and S.~Rigolin,
``Comprehensive ALP Searches in Meson Decays,''
(2025), [arXiv:2507.19578 [hep-ph]].

\bibitem{Bigazzi:2019hav}
F.~Bigazzi, A.~L.~Cotrone, M.~J\"arvinen, and E.~Kiritsis,
``Nonderivative Axionic Couplings to Nucleons at large and small $N$,''
JHEP \textbf{01}, 100 (2020).

\bibitem{Vonk:2021sit}
T.~Vonk, F.-K.~Guo, and U.-G.~Mei{\ss}ner,
``The axion-baryon coupling in SU(3) heavy baryon chiral perturbation theory,''
JHEP \textbf{08}, 024 (2021).

\bibitem{Cao:2024cym}
X.-H.~Cao and Z.-H.~Guo,
``Comprehensive study of axion photoproduction off the nucleon in chiral effective field theory,''
Phys. Rev. D \textbf{110}, 095025 (2024).

\bibitem{Gao:2022xqz}
R.~Gao, Z.-H.~Guo, J.~A.~Oller, and H.-Q.~Zhou,
``Axion-meson mixing in light of recent lattice $\eta$--$\eta'$ simulations and their two-photon couplings within U(3) chiral theory,''
JHEP \textbf{04}, 022 (2023).

\bibitem{Gan:2021?}
L.~Gan, B.~Kubis, E.~Passemar, and S.~Tulin,
``Precision tests of fundamental physics with $\eta$ and $\eta'$ mesons,''
Phys. Rept. \textbf{945}, 1--105 (2022).

\bibitem{Gao:2024vkw}
R.~Gao, J.~Hao, C.-G.~Duan, Z.-H.~Guo, J.~A.~Oller, and H.-Q.~Zhou,
``Isospin-breaking contribution to the model-independent axion-photon-photon coupling in U(3) chiral theory,''
Eur. Phys. J. C \textbf{85}, 97 (2025).

\bibitem{Feruglio:2025xvc}
F.~Feruglio, G.~Levati and R.~Ziegler,
``On the decay of a light spinless particle into two photons,''
JHEP \textbf{05} (2026), 307.

\bibitem{DiLuzio:2023cuk}
L.~Di Luzio, G.~Levati, and P.~Paradisi,
``The chiral Lagrangian of CP-violating axion-like particles,''
JHEP \textbf{02}, 020 (2024).

\bibitem{Goudzovski:2022vbt}
E.~Goudzovski et al.,
``New physics searches at kaon and hyperon factories,''
Rept. Prog. Phys. \textbf{86}, 016201 (2023).

\bibitem{REDTOP:2022slw}
J.~Elam et al.,
``The REDTOP experiment: Rare $\eta/\eta'$ Decays To Probe New Physics,''
(2022), 
[arXiv:2203.07651 [hep-ex]].

\bibitem{REDTOP:2026joh}
C.~Gatto et al.,
``Hidden-sectors search and probe of discrete symmetries at the REDTOP experiment,''(2026),
[arXiv:2606.12158 [hep-ex]].

\bibitem{NA62:2021zjw}
E.~Cortina Gil et al.,
``Measurement of the very rare $K^+\to\pi^+\nu\bar{\nu}$ decay,''
JHEP \textbf{06}, 093 (2021).

\bibitem{NA64:2020qwq}
D.~Banerjee et al.,
``Search for Axionlike and Scalar Particles with the NA64 Experiment,''
Phys. Rev. Lett. \textbf{125}, 081801 (2020).

\bibitem{KOTO:2020prk}
J.~K.~Ahn et al.,
``Study of the $K_L\to\pi^0\nu\bar{\nu}$ Decay at the J-PARC KOTO Experiment,''
Phys. Rev. Lett. \textbf{126}, 121801 (2021).

\bibitem{BESIII:2024hdv}
M.~Ablikim et al.,
``Search for diphoton decays of an axionlike particle in radiative $J/\psi$ decays,''
Phys. Rev. D \textbf{110}, L031101 (2024).

\bibitem{BESIII:2024awu}
M.~Ablikim et al.,
``Study of $\eta'\to\pi^+\pi^-\ell^+\ell^-$ decays at BESIII,''
JHEP \textbf{07}, 135 (2024).

\bibitem{Zhang:2026pyv}
J.~Zhang, J.~Fu, and H.-B.~Li,
``Searches for New Physics Beyond the Standard Model in Hyperon Sector,''
Chin. Phys. Lett. \textbf{43}, 060201 (2026).

\bibitem{Zielinski:2026lqa}
M.~Zieli\'nski, K.~Pro\'sci\'nski, and P.~Salabura,
``Search for the Axion-Like-Particles in the $\eta\to\pi^+\pi^-e^+e^-$ decay with HADES detector,''
PoS HADRON2025, 023 (2026).

\bibitem{Belle-II:2020jti}
F.~Abudin\'en et al.,
``Search for Axion-Like Particles produced in $e^+e^-$ collisions at Belle II,''
Phys. Rev. Lett. \textbf{125}, 161806 (2020).

\bibitem{MicroBooNE:2021sov}
P.~Abratenko et al.,
``Search for a Higgs Portal Scalar Decaying to Electron-Positron Pairs in the MicroBooNE Detector,''
Phys. Rev. Lett. \textbf{127}, 151803 (2021).

\bibitem{BaBar:2021ich}
J.~P.~Lees et al.,
``Search for an Axionlike Particle in $B$ Meson Decays,''
Phys. Rev. Lett. \textbf{128}, 131802 (2022).

\bibitem{Leutwyler:1996sa}
H.~Leutwyler,
``Bounds on the light quark masses,''
Phys. Lett. B \textbf{374}, 163--168 (1996).

\bibitem{Herrera-Siklody:1996tqr}
P.~Herrera-Siklody, J.~I.~Latorre, P.~Pascual, and J.~Taron,
``Chiral effective Lagrangian in the large $N_c$ limit: The Nonet case,''
Nucl. Phys. B \textbf{497}, 345--386 (1997).

\bibitem{Kaiser:2000gs}
R.~Kaiser and H.~Leutwyler,
``Large $N_c$ in chiral perturbation theory,''
Eur. Phys. J. C \textbf{17}, 623--649 (2000).

\bibitem{Ecker:2010nc}
G.~Ecker, P.~Masjuan, and H.~Neufeld,
``Chiral extrapolation and determination of low-energy constants from lattice data,''
Phys. Lett. B \textbf{692}, 184--188 (2010).

\bibitem{Ecker:2013pba}
G.~Ecker, P.~Masjuan, and H.~Neufeld,
``Approximating chiral SU(3) amplitudes,''
Eur. Phys. J. C \textbf{74}, 2748 (2014).

\bibitem{RBC:2010qam}
Y.~Aoki et al.,
``Continuum Limit Physics from 2+1 Flavor Domain Wall QCD,''
Phys. Rev. D \textbf{83}, 074508 (2011).

\bibitem{RBC:2012cbl}
R.~Arthur et al.,
``Domain Wall QCD with Near-Physical Pions,''
Phys. Rev. D \textbf{87}, 094514 (2013).

\end{thebibliography}

\end{document}